\documentclass[review]{elsarticle}

\usepackage[hyphens]{url}
\usepackage{graphicx}
\usepackage{booktabs}
\usepackage[T1]{fontenc}
\usepackage{lmodern}
\usepackage{caption}
\usepackage{subfig}
\usepackage{amssymb, amsmath}
\usepackage[inline]{enumitem}
\usepackage{float}
\usepackage{tabularx}
\usepackage[dvipsnames, table]{xcolor}
\usepackage{ifxetex, ifluatex}
\usepackage{fixltx2e}
\usepackage[unicode=true, colorlinks]{hyperref}
\usepackage{cleveref}
\usepackage{tabu}
\usepackage{mathpazo}

\usepackage{easyReview}

\IfFileExists{upquote.sty}{\usepackage{upquote}}{}

\ifnum 0\ifxetex 1\fi\ifluatex 1\fi=0 
  \usepackage[utf8]{inputenc}

\else 
  \usepackage{fontspec}
  \ifxetex
    \usepackage{xltxtra,xunicode}
  \fi
  \defaultfontfeatures{Mapping=tex-text,Scale=MatchLowercase}

\fi

\IfFileExists{microtype.sty}{\usepackage{microtype}}{}

\usepackage[margin=1in]{geometry}

\usepackage{color}
\usepackage{fancyvrb}

\DefineVerbatimEnvironment{Highlighting}{Verbatim}{commandchars=\\\{\}}
\usepackage{framed}
\definecolor{shadecolor}{RGB}{248,248,248}
\newenvironment{Shaded}{\begin{snugshade}}{\end{snugshade}}
\newcommand{\KeywordTok}[1]{\textcolor[rgb]{0.13,0.29,0.53}{\textbf{#1}}}
\newcommand{\DataTypeTok}[1]{\textcolor[rgb]{0.13,0.29,0.53}{#1}}
\newcommand{\DecValTok}[1]{\textcolor[rgb]{0.00,0.00,0.81}{#1}}

\newcommand{\FloatTok}[1]{\textcolor[rgb]{0.00,0.00,0.81}{#1}}

\newcommand{\StringTok}[1]{\textcolor[rgb]{0.31,0.60,0.02}{#1}}

\newcommand{\NormalTok}[1]{#1}

\usepackage{longtable}

\usepackage{setspace}
\usepackage{lmodern}

\providecommand{\tightlist}{%
  \setlength{\itemsep}{0pt}\setlength{\parskip}{0pt}}
  
\AtBeginDocument{%
  \newcommand\myshade{80}
  \colorlet{mylinkcolor}{violet!\myshade!black}
  \colorlet{mycitecolor}{YellowOrange!\myshade!black}
  \colorlet{myurlcolor}{Aquamarine!\myshade!black}

  \hypersetup{
    breaklinks = true,
    bookmarks  = true,
    pdfauthor  = {},
    pdftitle   = {Comparison of Multi-response Prediction Methods},
    linkcolor  = mylinkcolor,
    citecolor  = mycitecolor,
    urlcolor   = myurlcolor,
    colorlinks = true,
  }
}

\biboptions{numbers,sort&compress}

\makeatletter
\providecommand{\doi}[1]{%
  \begingroup
    \let\bibinfo\@secondoftwo
    \urlstyle{tt}%
    \href{http://dx.doi.org/#1}{%
      \discretionary{}{}{}%
      \nolinkurl{#1}%
    }%
  \endgroup
}
\makeatother

\begin{document}
\begin{frontmatter}

  \title{Comparison of Multi-response Prediction Methods}
  
    \author[KBM]{Raju Rimal\corref{c1}}
   \ead{raju.rimal@nmbu.no} 
   \cortext[c1]{Corresponding Author}
    \author[KBM]{Trygve Almøy}
   \ead{trygve.almoy@nmbu.no} 
  
    \author[NMBU]{Solve Sæbø}
   \ead{solve.sabo@nmbu.no} 
  
      \address[KBM]{Faculty of Chemistry and Bioinformatics, Norwegian University of Life
Sciences, Ås, Norway}
    \address[NMBU]{Professor, Norwegian University of Life Sciences, Ås, Norway}
  
  \begin{abstract}
  While data science is battling to extract information from the enormous
  explosion of data, many estimators and algorithms are being developed
  for better prediction. Researchers and data scientists often introduce
  new methods and evaluate them based on various aspects of data. However,
  studies on the impact of/on a model with multiple response variables are
  limited. This study compares some newly-developed (envelope) and well-
  established (PLS, PCR) prediction methods based on real data and
  simulated data specifically designed by varying properties such as
  multicollinearity, the correlation between multiple responses and
  position of relevant principal components of predictors. This study aims
  to give some insight into these methods and help the researcher to
  understand and use them in further studies.
  \end{abstract}

\end{frontmatter}

\section{Introduction}\label{introduction}

The prediction has been an essential component of modern data science,
whether in the discipline of statistical analysis or machine learning.
Modern technology has facilitated a massive explosion of data however,
such data often contain irrelevant information that consequently makes
prediction difficult. Researchers are devising new methods and
algorithms in order to extract information to create robust predictive
models. Such models mostly contain predictor variables that are directly
or indirectly correlated with other predictor variables. In addition,
studies often consist of many response variables correlated with each
other. These interlinked relationships influence any study, whether it
is predictive modelling or inference.

Modern inter-disciplinary research fields such as chemometrics,
econometrics and bioinformatics handle multi-response models
extensively. This paper attempts to compare some multivariate prediction
methods based on their prediction performance on linear model data with
specific properties. The properties include the correlation between
response variables, the correlation between predictor variables, number
of predictor variables and the position of relevant predictor
components. These properties are discussed more in the
\protect\hyperlink{experimental-design}{Experimental Design} section.
Among others, \citet{saebo2015simrel} and \citet{Alm_y_1996} have
conducted a similar comparison in the single response setting. In
addition, \citet{Rimal2018} have also conducted a basic comparison of
some prediction methods and their interaction with the data properties
of a multi-response model. The main aim of this paper is to present a
comprehensive comparison of contemporary prediction methods such as
simultaneous envelope estimation (Senv) \citep{cook2015simultaneous} and
envelope estimation in predictor space (Xenv) \citep{cook2010envelope}
with customary prediction methods such as Principal Component Regression
(PCR), Partial Least Squares Regression (PLS) using simulated dataset
with controlled properties. In the case of PLS, we have used PLS1 which
fits individual response separately and PLS2 which fits all the
responses together. Experimental design and the methods under comparison
are discussed further, followed by a brief discussion of the strategy
behind the data simulation.

\section{Simulation Model}\label{simulation-model}

Consider a model where the response vector \((\mathbf{y})\) with \(m\)
elements and predictor vector \((\mathbf{x})\) with \(p\) elements
follow a multivariate normal distribution as follows,

\begin{equation}
  \begin{bmatrix}
    \mathbf{y} \\ \mathbf{x}
  \end{bmatrix} \sim \mathcal{N}
  \left(
    \begin{bmatrix}
      \boldsymbol{\mu}_y \\
      \boldsymbol{\mu}_x
    \end{bmatrix},
    \begin{bmatrix}
    \boldsymbol{\Sigma}_{yy} & \boldsymbol{\Sigma}_{yx} \\
    \boldsymbol{\Sigma}_{xy} & \boldsymbol{\Sigma}_{xx}
    \end{bmatrix}
  \right)
  \label{eq:model-1}
\end{equation}

where, \(\boldsymbol{\Sigma}_{xx}\) and \(\boldsymbol{\Sigma}_{yy}\) are
the variance-covariance matrices of \(\mathbf{x}\) and \(\mathbf{y}\),
respectively, \(\boldsymbol{\Sigma}_{xy}\) is the covariance between
\(\mathbf{x}\) and \(\mathbf{y}\) and \(\boldsymbol{\mu}_x\) and
\(\boldsymbol{\mu}_y\) are mean vectors of \(\mathbf{x}\) and
\(\mathbf{y}\), respectively. A linear model based on \eqref{eq:model-1}
is,

\begin{equation}
\mathbf{y} = \boldsymbol{\mu}_y + 
  \boldsymbol{\beta}^t(\mathbf{x} - \boldsymbol{\mu_x}) + 
  \boldsymbol{\epsilon}
\label{eq:reg-model-1}
\end{equation}

where, \(\underset{m\times p}{\boldsymbol{\beta}^t}\) is a matrix of
regression coefficients and \(\boldsymbol{\epsilon}\) is an error term
such that
\(\boldsymbol{\epsilon} \sim \mathcal{N}(0, \boldsymbol{\Sigma}_{y|x})\).
Here,
\(\boldsymbol{\beta}^t = \mathbf{\Sigma}_{yx}\mathbf{\Sigma}_{xx}^{-1}\)
and
\(\boldsymbol{\Sigma}_{y|x} = \boldsymbol{\Sigma}_{yy} - \boldsymbol{\Sigma}_{yx}\boldsymbol{\Sigma}_{xx}^{-1}\boldsymbol{\Sigma}_{xy}\)

\begin{figure}

{\centering \includegraphics[width=0.8\linewidth]{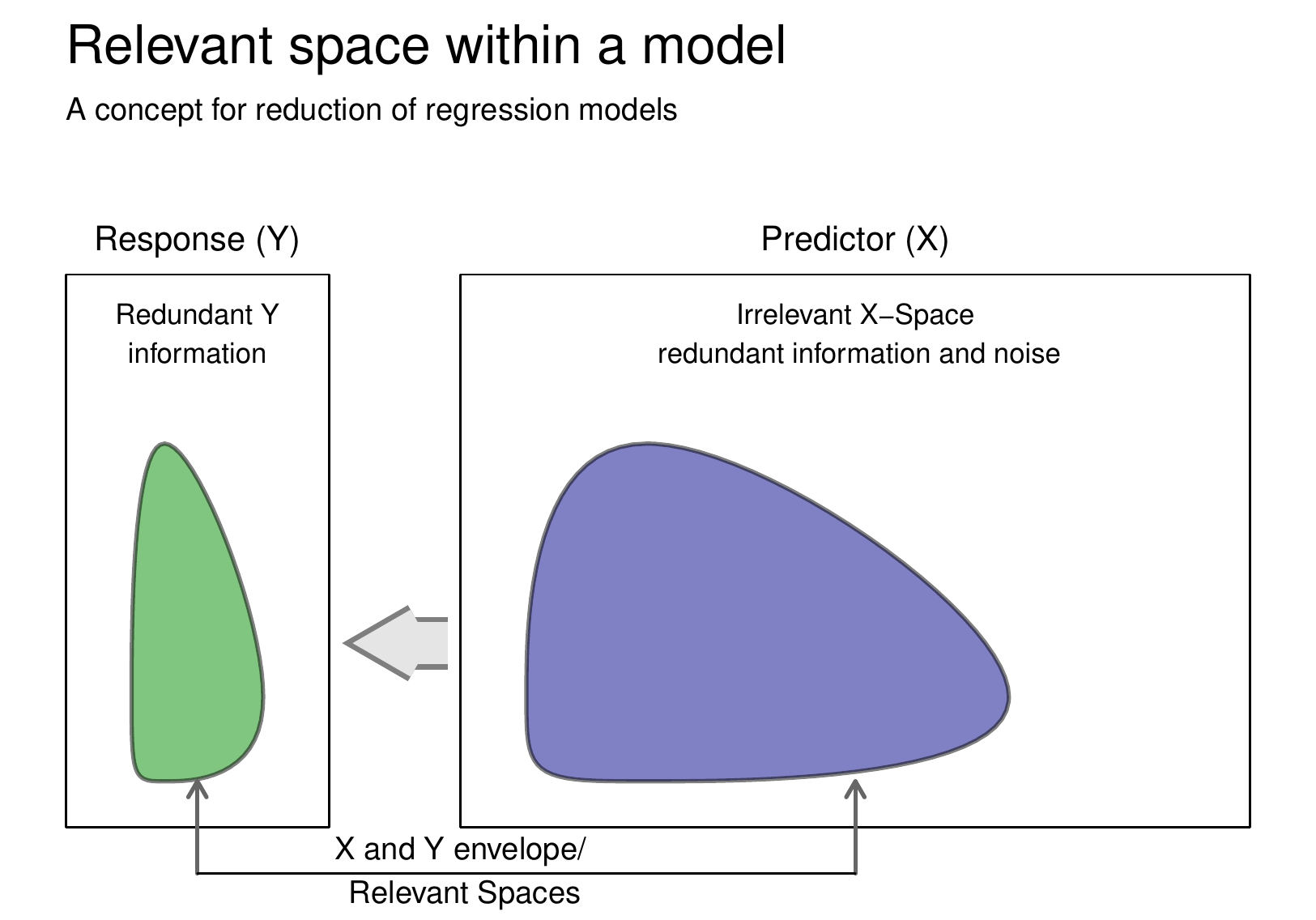} 

}

\caption{Relevant space in a regression model}\label{fig:relevant-space}
\end{figure}

In a model like \eqref{eq:reg-model-1}, we assume that the variation in
response \(\mathbf{y}\) is partly explained by the predictor
\(\mathbf{x}\). However, in many situations, only a subspace of the
predictor space is relevant for the variation in the response
\(\mathbf{y}\). This space can be referred to as the relevant space of
\(\mathbf{x}\) and the rest as irrelevant space. In a similar way, for a
certain model, we can assume that a subspace in the response space
exists and contains the information that the relevant space in predictor
can explain (Figure-\ref{fig:relevant-space}). \citet{cook2010envelope}
and \citet{cook2015simultaneous} have referred to the relevant space as
material space and the irrelevant space as immaterial space.

With an orthogonal transformation of \(\mathbf{y}\) and \(\mathbf{x}\)
to latent variables \(\mathbf{w}\) and \(\mathbf{z}\), respectively, by
\(\mathbf{w=Qy}\) and \(\mathbf{z = Rx}\), where \(\mathbf{Q}\) and
\(\mathbf{R}\) are orthogonal rotation matrices, an equivalent model to
\eqref{eq:model-1} in terms of the latent variables can be written as,

\begin{equation}
  \begin{bmatrix}
    \mathbf{w} \\ \mathbf{z}
  \end{bmatrix} \sim \mathcal{N}
  \left(
    \begin{bmatrix}
      \boldsymbol{\mu}_w \\
      \boldsymbol{\mu}_z
    \end{bmatrix},
    \begin{bmatrix}
    \boldsymbol{\Sigma}_{ww} & \boldsymbol{\Sigma}_{wz} \\
    \boldsymbol{\Sigma}_{zw} & \boldsymbol{\Sigma}_{zz}
    \end{bmatrix}
  \right)
  \label{eq:model-2}
\end{equation}

where, \(\boldsymbol{\Sigma}_{ww}\) and \(\boldsymbol{\Sigma}_{zz}\) are
the variance-covariance matrices of \(\mathbf{w}\) and \(\mathbf{z}\),
respectively. \(\boldsymbol{\Sigma}_{zw}\) is the covariance between
\(\mathbf{z}\) and \(\mathbf{w}\). \(\boldsymbol{\mu}_w\) and
\(\boldsymbol{\mu}_z\) are the mean vector of \(\mathbf{z}\) and
\(\mathbf{w}\) respectively.

Here, the elements of \(\mathbf{w}\) and \(\mathbf{z}\) are the
principal components of responses and predictors, which will
respectively be referred to respectively as ``response components'' and
``predictor components''. The column vectors of respective rotation
matrices \(\mathbf{Q}\) and \(\mathbf{R}\) are the eigenvectors
corresponding to these principal components. We can write a linear model
based on \eqref{eq:model-2} as,

\begin{equation}
\mathbf{w} = \boldsymbol{\mu}_w + \boldsymbol{\alpha}^t(\mathbf{z} - \boldsymbol{\mu_z}) + \boldsymbol{\tau}
\label{eq:reg-model-2}
\end{equation}

where, \(\underset{m\times p}{\boldsymbol{\alpha}^t}\) is a matrix of
regression coefficients and \(\boldsymbol{\tau}\) is an error term such
that
\(\boldsymbol{\tau} \sim \mathcal{N}(0, \boldsymbol{\Sigma}_{w|z})\).

Following the concept of relevant space, a subset of predictor
components can be imagined to span the predictor space. These components
can be regarded as relevant predictor components. \citet{Naes1985}
introduced the concept of relevant components which was explored further
by \citet{helland1990partial}, \citet{naes1993relevant},
\citet{Helland1994b} and \citet{Helland2000}. The corresponding
eigenvectors were referred to as relevant eigenvectors. A similar logic
is introduced by \citet{cook2010envelope} and later by
\citet{cook2013envelopes} as an envelope which is the space spanned by
the relevant eigenvectors \citep[pp.~101]{cook2018envelope}.

In addition, various simulation studies have been performed with the
model based on the concept of relevant subspace. A simulation study by
\citet{Alm_y_1996} has used a single response simulation model based on
reduced regression and has compared some contemporary multivariate
estimators. In recent years \citet{helland2012near},
\citet{saebo2015simrel}, \citet{helland2016algorithms} and
\citet{Rimal2018} implemented similar simulation examples similar to
those we are discussing in this study. This paper, however, presents an
elaborate comparison of the prediction using multi-response simulated
linear model data. The properties of the simulated data are varied
through different levels of simulation-parameters based on an
experimental design. \citet{Rimal2018} provide a detailed discussion of
the simulation model that we have adopted here. The following section
presents the estimators being compared in more detail.

\section{Prediction Methods}\label{prediction-methods}

Partial least squares regression (PLS) and Principal component
regression (PCR) have been used in many disciplines such as
chemometrics, econometrics, bioinformatics and machine learning, where
wide predictor matrices, i.e. \(p\) (number or predictors)
\textgreater{} \(n\) (number of observation) are common. These methods
are popular in multivariate analysis, especially for exploratory studies
and predictions. In recent years, a concept of envelope introduced by
\citet{Cook2007a} based on the reduction in the regression model was
implemented for the development of different estimators. This study
compares these prediction methods based on their prediction performance
on data simulated with different controlled properties.

\begin{description}
\tightlist
\item[\emph{Principal Components Regression (PCR):}]
Principal components are the linear combinations of predictor variables
such that the transformation makes the new variables uncorrelated. In
addition, the variation of the original dataset captured by the new
variables is sorted in descending order. In other words, each successive
component captures maximum variation left by the preceding components in
predictor variables \citep{Jolliffe2002}. Principal components
regression uses these principal components as a new predictor to explain
the variation in the response.
\item[\emph{Partial Least Squares (PLS):}]
Two variants of PLS: PLS1 and PLS2 are used for comparison. The first
one considers individual response variables separately, i.e.~each
response is predicted with a single response model, while the latter
considers all response variables together. In PLS regression, the
components are determined so as to maximize a covariance between
response and predictors \citep{DeJong1993}. R-package \texttt{pls}
\citep{pls2018} is used for both PCR and PLS methods.
\item[\emph{Envelopes:}]
The envelope, introduced by \citet{Cook2007a}, was first used to define
response envelope \citep{cook2010envelope} as the smallest subspace in
the response space so that the span of regression coefficients lies in
that space. Since a multivariate linear regression model contains
relevant (material) and irrelevant (immaterial) variation in both
response and predictor, the relevant part provides information, while
the irrelevant part increases the estimative variation. The concept of
the envelope uses the relevant part for estimation while excluding the
irrelevant part consequently increasing the efficiency of the model
\citep{cook2016algorithms}.

The concept was later extended to the predictor space, where the
predictor envelope was defined \citep{cook2013envelopes}. Further
\citet{cook2015simultaneous} used envelopes for joint reduction of the
responses and predictors and argued that this produced efficiency gains
that were greater than those derived by using individual envelopes for
either the responses or the predictors separately. All the variants of
envelope estimations are based on maximum likelihood estimation. Here we
have used predictor envelope (Xenv) and simultaneous envelope (Senv) for
the comparison. R-package \texttt{Renvlp} \citep{env2018} is used for
both Xenv and Senv methods.
\end{description}

\subsection{Modification in envelope
estimation}\label{modification-in-envelope-estimation}

Since envelope estimators (Xenv and Senv) are based on maximum
likelihood estimation (MLE), it fails to estimate in the case of wide
matrices, i.e. \(p > n\). To incorporate these methods in our
comparison, we have used the principal components \((\mathbf{z})\) of
the predictor variables \((\mathbf{x})\) as predictors, using the
required number of components for capturing 97.5\% of the variation in
\(\mathbf{x}\) for the designs where \(p > n\). The new set of variables
\(\mathbf{z}\) were used for envelope estimation. The regression
coefficients \((\hat{\boldsymbol{\alpha}})\) corresponding to these new
variables \(\mathbf{z}\) were transformed back to obtain coefficients
for each predictor variable as,
\[\hat{\boldsymbol{\beta}} = \mathbf{e}_k\hat{\boldsymbol{\alpha}_k}\]
where \(\mathbf{e}_k\) is a matrix of eigenvectors with the first \(k\)
number of components. Only simultaneous envelope allows to specify the
dimension of response envelope and all the simulation is based on a
single latent dimension of response, so it is fixed at two in the
simulation study. In the case of Senv, when the envelope dimension for
response is the same as the number of responses, it degenerates to the
Xenv method and if the envelope dimension for the predictor is the same
as the number of predictors, it degenerates to the standard multivariate
linear regression \citep{env2018}.

\hypertarget{experimental-design}{\section{Experimental
Design}\label{experimental-design}}

This study compares prediction methods based on their prediction
ability. Data with specific properties are simulated, some of which are
easier to predict than others. These data are simulated using the
R-package \texttt{simrel}, which is discussed in \citet{saebo2015simrel}
and \citet{Rimal2018}. Here we have used four different factors to vary
the property of the data: a) Number of predictors (\texttt{p}), b)
Multicollinearity in predictor variables (\texttt{gamma}), c)
Correlation in response variables (\texttt{eta}) and d) position of
predictor components relevant for the response (\texttt{relpos}). Using
two levels of \texttt{p}, \texttt{gamma} and \texttt{relpos} and four
levels of \texttt{eta}, 32 set of distinct properties are designed for
the simulation.

\begin{description}
\item[\textbf{Number of predictors:}]
To observe the performance of the methods on tall and wide predictor
matrices, 20 and 250 predictor variables are simulated with the number
of observations fixed at 100. Parameter \texttt{p} controls these
properties in the \texttt{simrel} function.
\item[\textbf{Multicollinearity in predictor variables:}]
Highly collinear predictors can be explained completely by a few
components. The parameter \texttt{gamma} (\(\gamma\)) in \texttt{simrel}
controls decline in the eigenvalues of the predictor variables as
\eqref{eq:gamma}.

\begin{equation}
  \lambda_i = e^{-\gamma(i - 1)}, \gamma > 0 \text{ and } i = 1, 2, \ldots, p
  \label{eq:gamma}
\end{equation}

Here, \(\lambda_i, i = 1, 2, \ldots p\) are eigenvalues of the predictor
variables. We have used 0.2 and 0.9 as different levels of
\texttt{gamma}. The higher the value of gamma, the higher the
multicollinearity will be, and vice versa.
\item[\textbf{Correlation in response variables:}]
Correlation among response variables has been explored to a lesser
extent. Here we have tried to explore that part with four levels of
correlation in the response variables. We have used the \texttt{eta}
(\(\eta\)) parameter of \texttt{simrel} for controlling the decline in
eigenvalues corresponding to the response variables as \eqref{eq:eta}.

\begin{equation}
  \kappa_j = e^{-\eta(j - 1)}, \eta > 0 \text{ and } j = 1, 2, \ldots, m
  \label{eq:eta}
\end{equation}

Here, \(\kappa_j, i = 1, 2, \ldots m\) are the eigenvalues of the
response variables and \texttt{m} is the number of response variables.
We have used 0, 0.4, 0.8 and 1.2 as different levels of \texttt{eta}.
The larger the value of eta, the larger will be the correlation will be
between response variables and vice versa.
\item[\textbf{Position of predictor components relevant to the
response:}]
The principal components of the predictors are ordered. The first
principal component captures most of the variation in the predictors.
The second captures most of the remainder left by the first principal
component and so on. In highly collinear predictors, the variation
captured by the first few components is relatively high. However, if
those components are not relevant for the response, prediction becomes
difficult \citep{Helland1994b}. Here, two levels of the positions of
these relevant components are used as 1, 2, 3, 4 and 5, 6, 7, 8.
\end{description}

\begin{figure}
\includegraphics[width=1\linewidth]{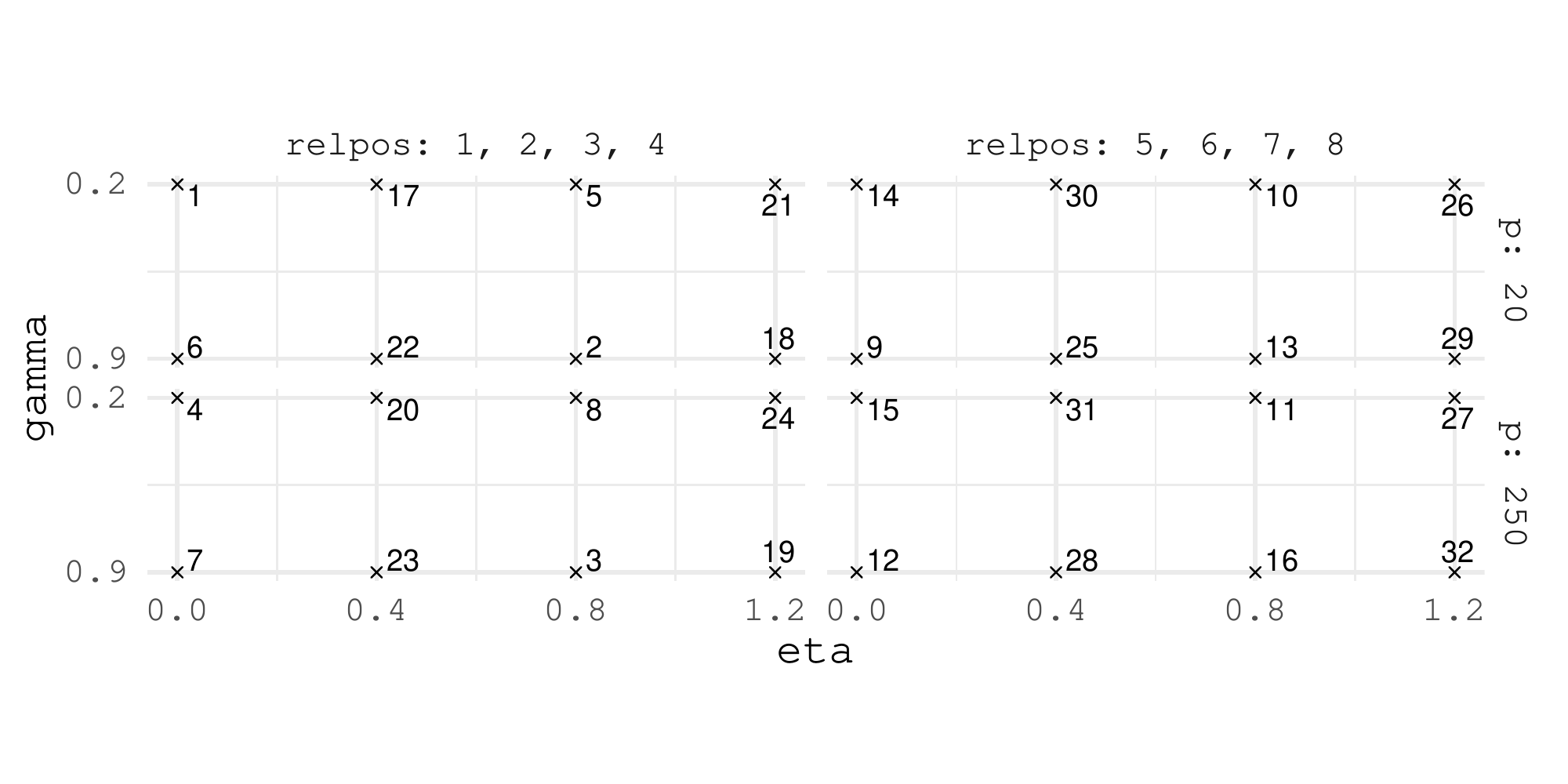} \caption{Experimental Design of simulation parameters. Each point represents a unique data property.}\label{fig:design-plot}
\end{figure}

Moreover, a complete factorial design from the levels of the above
parameters gave us 32 designs. Each design is associated with a dataset
having unique properties. Figure\textasciitilde{}\ref{fig:design-plot},
shows all the designs. For each design and prediction method, 50
datasets were simulated as replicates. In total, there were
\(5 \times 32 \times 50\), i.e.~8000 simulated datasets.

\begin{description}
\tightlist
\item[\textbf{Common parameters:}]
Each dataset was simulated with \(n = 100\) number of observation and
\(m = 4\) response variables. Furthermore, the coefficient of
determination corresponding to each response components in all the
designs is set to 0.8. In addition, we have assumed that there is only
one informative response component. Hence, the informative response
component is rotated orthogonally together with three uninformative
response components to generate four response variables. This spreads
out the information in all simulated response variables. For further
details on the simulation tool, see \citep{Rimal2018}.
\end{description}

An example of simulation parameters for the first design is as follows:

\begin{Shaded}
\begin{Highlighting}[]
\KeywordTok{simrel}\NormalTok{(}
    \DataTypeTok{n       =} \DecValTok{100}\NormalTok{,                 ## Training samples}
    \DataTypeTok{p       =} \DecValTok{20}\NormalTok{,                  ## Predictors}
    \DataTypeTok{m       =} \DecValTok{4}\NormalTok{,                   ## Responses}
    \DataTypeTok{q       =} \DecValTok{20}\NormalTok{,                  ## Relevant predictors}
    \DataTypeTok{relpos  =} \KeywordTok{list}\NormalTok{(}\KeywordTok{c}\NormalTok{(}\DecValTok{1}\NormalTok{, }\DecValTok{2}\NormalTok{, }\DecValTok{3}\NormalTok{, }\DecValTok{4}\NormalTok{)), ## Relevant predictor components index}
    \DataTypeTok{eta     =} \DecValTok{0}\NormalTok{,                   ## Decay factor of response eigenvalues}
    \DataTypeTok{gamma   =} \FloatTok{0.2}\NormalTok{,                 ## Decay factor of predictor eigenvalues}
    \DataTypeTok{R2      =} \FloatTok{0.8}\NormalTok{,                 ## Coefficient of determination}
    \DataTypeTok{ypos    =} \KeywordTok{list}\NormalTok{(}\KeywordTok{c}\NormalTok{(}\DecValTok{1}\NormalTok{, }\DecValTok{2}\NormalTok{, }\DecValTok{3}\NormalTok{, }\DecValTok{4}\NormalTok{)),}
    \DataTypeTok{type    =} \StringTok{"multivariate"}
\NormalTok{)}
\end{Highlighting}
\end{Shaded}

\begin{figure}
\includegraphics[width=1\linewidth]{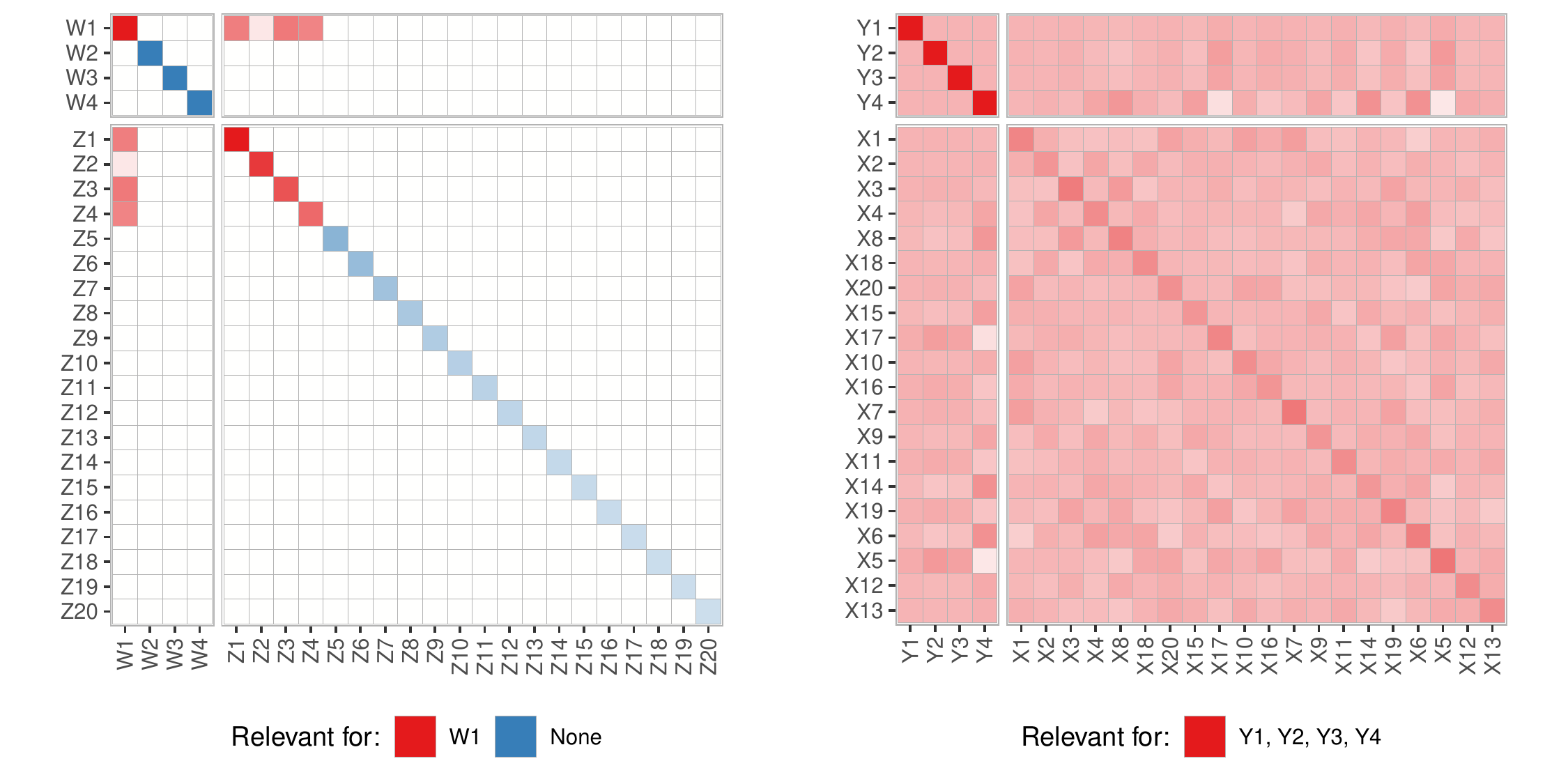} \caption{(left) Covariance structure of latent components (right) Covariance structure of predictor and response}\label{fig:cov-plot-1}
\end{figure}

\begin{figure}
\includegraphics[width=1\linewidth]{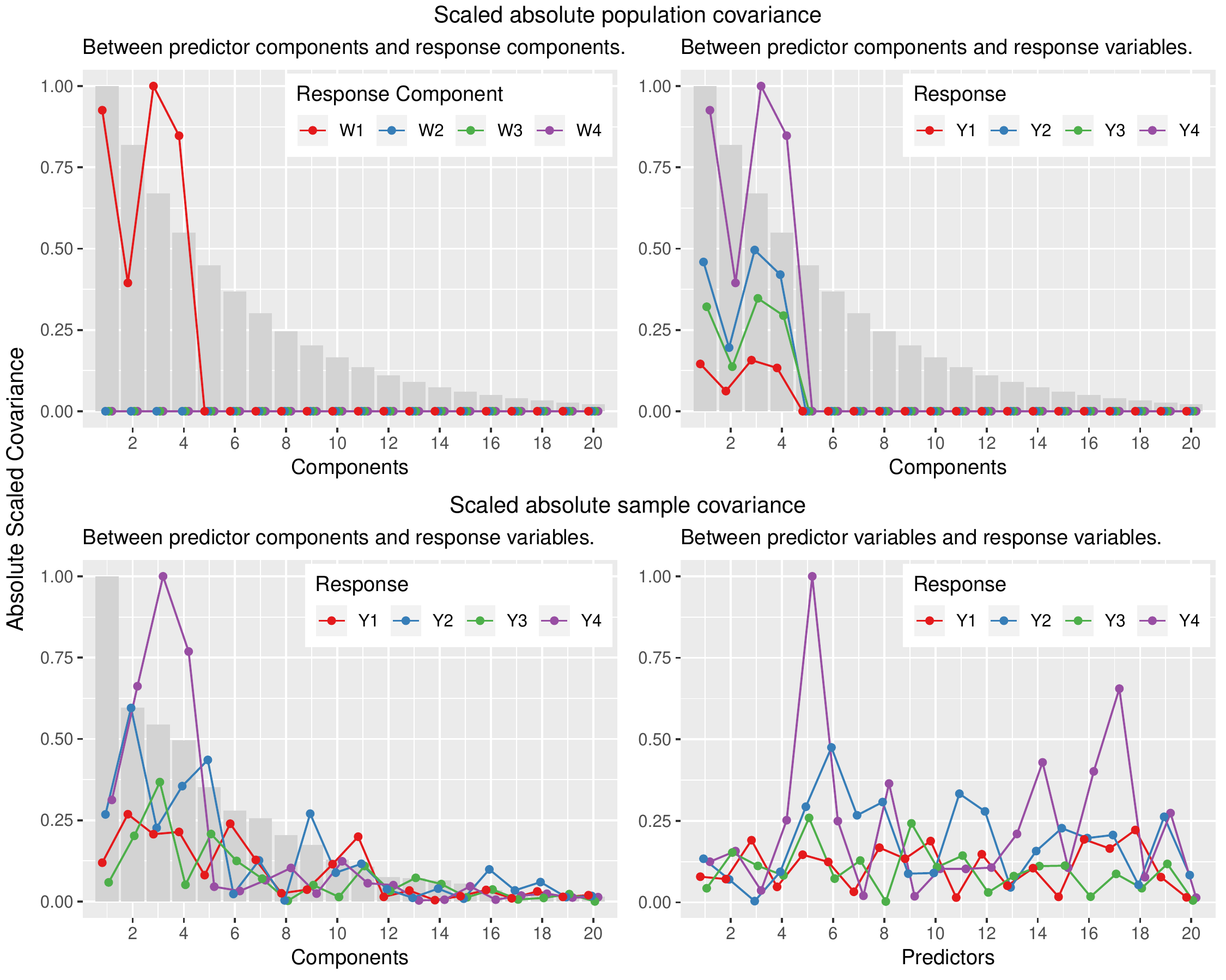} \caption{Expected Scaled absolute covariance between predictor
components and response components (top left). Expected Scaled absolute
covariance between predictor components and response variables (top
right). Sample scaled absolute covariance between predictor components
and response variables (bottom left). Sample scaled absolute covariance
between predictor variables and response variables (bottom right). The
bar graph in the background represents eigenvalues corresponding to each
component in the population (top plots) and in the sample (bottom
plots). One can compare the top-right plot (true covariance of the
population) with bottom-left (covariance in the simulated data) which
shows a similar pattern for different components.}\label{fig:est-cov-plot}
\end{figure}

The covariance structure of the data simulated with this design in the
Figure \ref{fig:cov-plot-1} shows that the predictor components at
positions 1, 2, 3 and 4 are relevant for the first response component.
After the rotation with an orthogonal rotation matrix, all predictor
variables are somewhat relevant for all response variables, satisfying
other desired properties such as multicollinearity and coefficient of
determination. For the same design, Figure \ref{fig:est-cov-plot} (top
left) shows that the predictor components 1, 2, 3 and 4 are relevant for
the first response component. All other predictor components are
irrelevant and all other response components are uninformative. However,
due to orthogonal rotation of the informative response component
together with uninformative response components, all response variables
in the population have similar covariance with the relevant predictor
components (Figure \ref{fig:est-cov-plot} (top right)). The sample
covariances between the predictor components and predictor variables
with response variables are shown in Figure \ref{fig:est-cov-plot}
(bottom left) and (bottom right) respectively.

A similar description can be made for all 32 designs, where each of the
designs holds the properties of the data they simulate. These data are
used by the prediction methods discussed in the previous section. Each
prediction method is given independently simulated datasets in order to
give them an equal opportunity to capture the dynamics in the data.

\section{Basis of comparison}\label{basis-of-comparison}

This study focuses mainly on the prediction performance of the methods
with an emphasis specifically on the interaction between the properties
of the data controlled by the simulation parameters and the prediction
methods. The prediction performance is measured based on the following:

\begin{enumerate}
\def\labelenumi{\alph{enumi})}
\tightlist
\item
  The average prediction error that a method can give using an arbitrary
  number of components and
\item
  The average number of components used by the method to give the
  minimum prediction error
\end{enumerate}

Let us define,

\begin{equation}
\mathcal{PE}_{ijkl} = \frac{1}{\sigma_{y_{ij}|x}^2}
  \mathsf{E}{\left[\left(\boldsymbol{\beta}_{ij} -
  \boldsymbol{\hat{\beta}_{ijkl}}\right)^t
  \left(\boldsymbol{\Sigma}_{xx}\right)_i
  \left(\boldsymbol{\beta}_{ij} - \boldsymbol{\hat{\beta}_{ijkl}}\right)\right]} + 1
\label{eq:pred-error}
\end{equation}

as a prediction error of response \(j = 1, \ldots 4\) for a given design
\(i=1, 2, \ldots 32\) and method
\(k=1(\text{PCR}), \ldots 5(\text{Senv})\) using \(l=0, \ldots 10\)
number of components. Here, \(\left(\boldsymbol{\Sigma}_{xx}\right)_i\)
is the true covariance matrix of the predictors, unique for a particular
design \(i\) and \(\sigma_{y_j\mid x}^2\) for response
\(j = 1, \ldots m\) is the true model error. Here prediction error is
scaled by the true model error to remove the effects of influencing
residual variances. Since both the expectation and the variance of
\(\hat{\boldsymbol{\beta}}\) are unknown, the prediction error is
estimated using data from 50 replications as follows,

\begin{equation}
\widehat{\mathcal{PE}_{ijkl}} = \frac{1}{\sigma_{y_{ij}|x}^2}
  \sum_{r=0}^{50}{\left[\left(\boldsymbol{\beta}_{ij} -
  \boldsymbol{\hat{\beta}_{ijklr}}\right)^t
  \left(\boldsymbol{\Sigma}_{xx}\right)_i
  \left(\boldsymbol{\beta}_{ij} - \boldsymbol{\hat{\beta}_{ijklr}}\right)\right]} + 1
\label{eq:estimated-pred-error}
\end{equation}

where \(\widehat{\mathcal{PE}_{ijkl}}\) is the estimated prediction
error averaged over \(r=50\) replicates.

The following section focuses on the data for the estimation of these
prediction errors that are used for the two models discussed above in a)
and b) of this section.

\section{Data Preparation}\label{data-preparation}

A dataset for estimating \eqref{eq:pred-error} is obtained from simulation
which contains a) five factors corresponding to simulation parameters,
b) prediction methods, c) number of components, d) replications and e)
prediction error for four responses. The prediction error is computed
using predictor components ranging from 0 to 10 for every 50 replicates
as,

\begin{equation*}
\left(\widehat{\mathcal{PE_\circ}}\right)_{ijklr} =
  \frac{1}{\sigma_{y_{ij}\mid x}^2}\left[
    \left(\boldsymbol{\beta}_{ij} - \hat{\boldsymbol{\beta}}_{ijklr}\right)^t
    \left(\boldsymbol{\Sigma}_{xx}\right)_{i}
    \left(\boldsymbol{\beta}_{ij} - \hat{\boldsymbol{\beta}}_{ijklr}\right)
  \right] + 1
\end{equation*}

Thus there are 32 (designs) \(\times\) 5 (methods) \(\times\) 11 (number
of components) \(\times\) 50 (replications), i.e.~88000 observations
corresponding to the response variables from \texttt{Y1} to \texttt{Y4}.

Since our discussions focus on the average minimum prediction error that
a method can obtain and the average number of components they use to get
the minimum prediction error in each replicates, the dataset discussed
above is summarized as constructing the following two smaller datasets.
Let us call them \emph{Error Dataset} and \emph{Component Dataset}.

\begin{description}
\item[\emph{Error Dataset}:]
For each prediction method, design and response, an average prediction
error is computed over all replicates for each component. Next, a
component that gives the minimum of this average prediction error is
selected, i.e.,

\begin{equation}
  l_\circ = \operatorname*{argmin}_{l}\left[\frac{1}{50}\sum_{i=1}^{50}{\left(\mathcal{PE}_\circ\right)_{ijklr}}\right]
  \label{eq:min-pred}
  \end{equation}

Using the component \(l_\circ\), a dataset of
\(\left(\mathcal{PE}_\circ\right)_{ijkl_\circ r}\) is used as the
\emph{Error Dataset}. Let \(\mathbf{u}_{(8000 \times 4)} = (u_j)\) for
\(j = 1, \ldots 4\) be the outcome variables measuring the prediction
error corresponding to the response number \(j\) in the context of this
dataset.
\item[\emph{Component Dataset}:]
The number of components that gives the minimum prediction error in each
replication is referred to as the \emph{Component Dataset}, i.e.,

\begin{equation}
  l_{\circ} = \operatorname*{argmin}_{l}\left[\mathcal{PE}_{ijklr}\right]
  \label{eq:min-comp}
  \end{equation}

Here \(l_\circ\) is the number of components that gives minimum
prediction error \(\left(\mathcal{PE}_\circ\right)_{ijklr}\) for design
\(i\), response \(j\), method \(k\) and replicate \(r\). Let
\(\mathbf{v}_{(8000 \times 4)} = (v_j)\) for \(j = 1, \ldots 4\) be the
outcome variables measuring the number of components used for minimum
prediction error corresponding to the response \(j\) in the context of
this dataset.
\end{description}

\section{Exploration}\label{exploration}

\begin{figure}[!htb]
\includegraphics[width=1\linewidth]{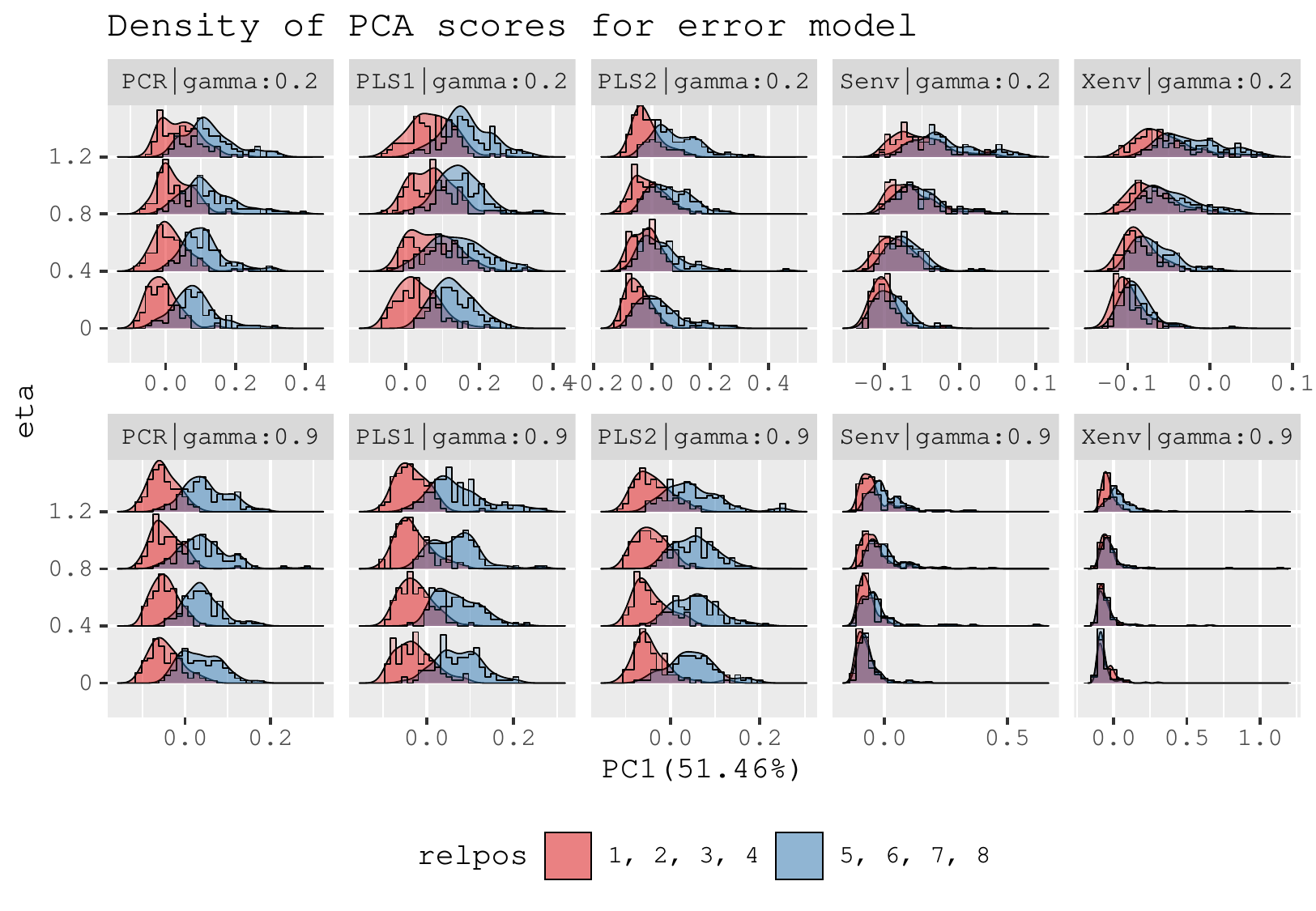} \caption{Scores density corresponding to first principal
component of \emph{error dataset} (\(\mathbf{u}\)) subdivided by
\texttt{methods}, \texttt{gamma} and \texttt{eta} and grouped by
\texttt{relpos}.}\label{fig:pred-pca-hist-mthd-gamma-relpos}
\end{figure}

This section explores the variation in the \emph{error dataset} and the
\emph{component dataset} for which we have used Principal Component
Analysis (PCA). Let \(\mathbf{t}_u\) and \(\mathbf{t}_v\) be the
principal component score sets corresponding to PCA run on the
\(\mathbf{u}\) and \(\mathbf{v}\) matrices respectively. The scores
density in Figure-\ref{fig:pred-pca-hist-mthd-gamma-relpos} corresponds
to the first principal component of \(\mathbf{u}\), i.e.~the first
column of \(\mathbf{t}_u\).

Since higher prediction errors correspond to high scores, the plot shows
that the PCR, PLS1 and PLS2 methods are influenced by the two levels of
the position of relevant predictor components. When the relevant
predictors are at positions 5, 6, 7, 8, the eigenvalues corresponding to
them are relatively smaller. This also suggests that PCR, PLS1 and PLS2
depend greatly on the position of the relevant components, and the
variation of these components affects their prediction performance.
However, the envelope methods appeared to be less influenced by
\texttt{relpos} in this regard.

In addition, the plot also shows that the effect of \texttt{gamma},
i.e., the level of multicollinearity, has a lesser effect when the
relevant predictors are at positions 1, 2, 3, 4. This indicates that the
methods are somewhat robust for handling collinear predictors.
Nevertheless, when the relevant predictors are at positions 5, 6, 7, 8,
high multicollinearity results in a small variance of these relevant
components and consequently yields poor prediction. This is in
accordance with the findings of \citet{Helland1994b}.

Furthermore, the density curves for PCR, PLS1 and PLS2 are similar for
different levels of \texttt{eta}, i.e., the factor controlling the
correlation between responses. However, the envelope models have been
shown to have distinct interactions between the positions of relevant
components (\texttt{relpos}) and \texttt{eta}. Here higher levels of
\texttt{eta} have yielded higher scores and clear separation between two
levels of \texttt{relpos}.

In the case of high multicollinearity, envelope methods have resulted in
some large outliers indicating that in some cases that the methods can
result in giving an unexpected prediction.

\begin{figure}[!htb]
\includegraphics[width=1\linewidth]{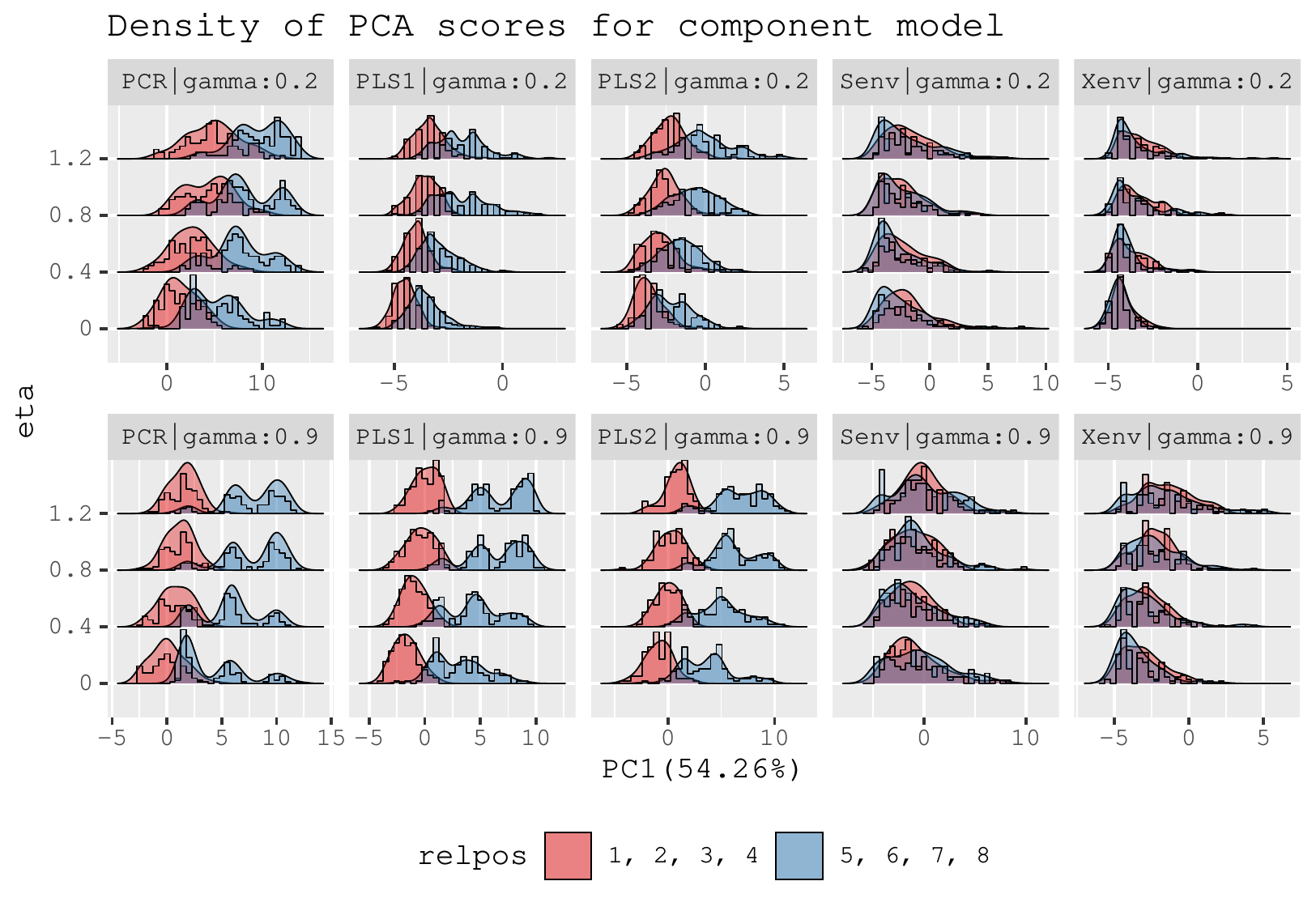} \caption{Score density corresponding to first principal component
of \emph{component dataset} (\(\mathbf{v}\)) subdivided by
\texttt{methods}, \texttt{gamma} and \texttt{eta} and grouped by
\texttt{relpos}.}\label{fig:comp-pca-hist-mthd-gamma-relpos}
\end{figure}

In Figure \ref{fig:comp-pca-hist-mthd-gamma-relpos}, the higher scores
suggest that methods have used a larger number of components to give
minimum prediction error. The plot also shows that the relevant
predictor components at 5, 6, 7, 8 give larger prediction errors than
those in positions 1, 2, 3, 4. The pattern is more distinct in large
multicollinearity cases and PCR and PLS methods. Both the envelope
methods have shown equally enhanced performance at both levels of
\texttt{relpos} and \texttt{gamma}. However, for data with low
multicollinearity (\(\gamma = 0.2\)), the envelope methods have used a
lesser number of components on average than in the high
multicollinearity cases to achieve minimum prediction error.

\section{Statistical Analysis}\label{statistical-analysis}

This section has modelled the \emph{error data} and the \emph{component
data} as a function of the simulation parameters to better understand
the connection between data properties and prediction methods using
multivariate analysis of variation (MANOVA).

Let us consider a model with third order interaction of the simulation
parameters (\texttt{p}, \texttt{gamma}, \texttt{eta} and
\texttt{relpos}) and \texttt{Methods} as in \eqref{eq:error-model} and
\eqref{eq:component-model} using datasets \(\mathbf{u}\) and
\(\mathbf{v}\), respectively. Let us refer them as the \emph{error
model} and the \emph{component model}.

\begin{description}
\item[\textbf{Error Model:}]
\begin{equation}\mathbf{u}_{abcdef} = \boldsymbol{\mu}_u +
  (\texttt{p}_a + \texttt{gamma}_b + \texttt{eta}_c +
\texttt{relpos}_d + \texttt{Methods}_e)^3 +
  \left(\boldsymbol{\varepsilon}_u\right)_{abcdef}
  \label{eq:error-model}
  \end{equation}
\item[\textbf{Component Model:}]
\begin{equation}\mathbf{v}_{abcdef} = \boldsymbol{\mu}_v +
  (\texttt{p}_a + \texttt{gamma}_b + \texttt{eta}_c +
\texttt{relpos}_d + \texttt{Methods}_e)^3 +
  \left(\boldsymbol{\varepsilon}_v\right)_{abcdef}
  \label{eq:component-model}
  \end{equation}
\end{description}

where, \(\mathbf{u}_{abcdef}\) is a vector of prediction errors in the
\emph{error model} and \(\mathbf{v}_{abcdef}\) is a vector of the number
of components used by a method to obtain minimum prediction error in the
\emph{component model}.

Although there are several test-statistics for MANOVA, all are
essentially equivalent for large samples \citep{johnson2018applied}.
Here we will use Pillai's trace statistic which is defined as,

\begin{equation}
\text{Pillai statistic} = \text{tr}\left[
\left(\mathbf{E} + \mathbf{H}\right)^{-1}\mathbf{H}
\right] = \sum_{i=1}^m{\frac{\nu_i}{1 + \nu_i}}
\label{eq:pillai}
\end{equation}

Here the matrix \(\mathbf{H}\) holds between-sum-of-squares and
sum-of-products for each of the predictors. The matrix \(\mathbf{E}\)
has a within the sum of squares and sum of products for each of the
predictors. \(\nu_i\) represents the eigenvalues corresponding to
\(\mathbf{E}^{-1}\mathbf{H}\) \citep{rencher2003methods}.

\begin{figure}
\includegraphics[width=1\linewidth]{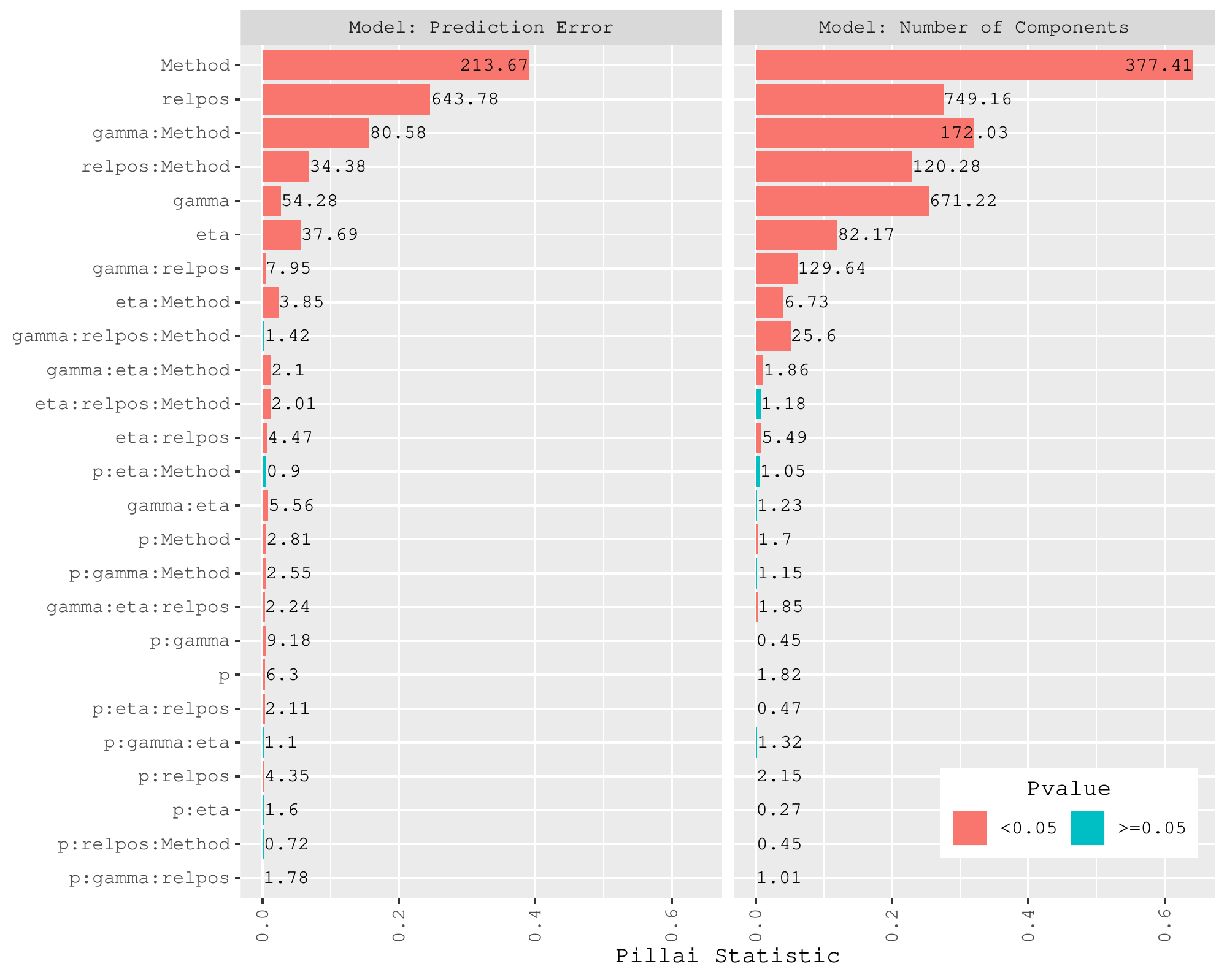} \caption{Pillai Statistic and F-value for the MANOVA model. The
bar represents the Pillai Statistic and the text labels are F-value for
the corresponding factor.}\label{fig:manova-plot}
\end{figure}

For both the models \eqref{eq:error-model} and \eqref{eq:component-model},
Pillai's trace statistic is used for accessing the effect of each factor
and returns an F-value for the strength of their significance. Figure
\ref{fig:manova-plot} plots the Pillai's trace statistics as bars with
corresponding F-values as text labels for both models.

\begin{description}
\tightlist
\item[Error Model:]
Figure \ref{fig:manova-plot} (left) shows the Pillai's trace statistic
for factors of the \emph{error model}. The main effect of
\texttt{Method} followed by \texttt{relpos}, \texttt{eta} and
\texttt{gamma} have largest influence on the model. A highly significant
two-factor interaction of \texttt{Method} with \texttt{eta} followed by
\texttt{relpos} and \texttt{gamma} clearly shows that methods perform
differently for different levels of these data properties. The
significant third order interaction between \texttt{Method},
\texttt{eta} and \texttt{gamma} suggests that the performance of a
method differs for a given level of multicollinearity and the
correlation between the responses. Since only some methods consider
modelling predictor and response together, the prediction is affected by
the level of correlation between the responses (\texttt{eta}) for a
given method.
\item[Component Model:]
Figure \ref{fig:manova-plot} (right) shows the Pillai's trace statistic
for factors of the \emph{component model}. As in the \emph{error model},
the main effects of the Method, \texttt{relpos}, \texttt{gamma} and
\texttt{eta} have a significantly large effect on the number of
components that a method has used to obtain minimum prediction error.
The two-factor interactions of \texttt{Method} with simulation
parameters are larger in this case. This shows that the Methods and
these interactions have a larger effect on the use of the number of
component than the prediction error itself. In addition, a similar
significant high third-order interaction as found in the \emph{error
model} is also observed in this model.
\end{description}

The following section will continue to explore the effects of different
levels of the factors in the case of these interactions.

\subsection{Effect Analysis of Error
Model}\label{effect-analysis-of-error-model}

\begin{figure}
\includegraphics[width=1\linewidth]{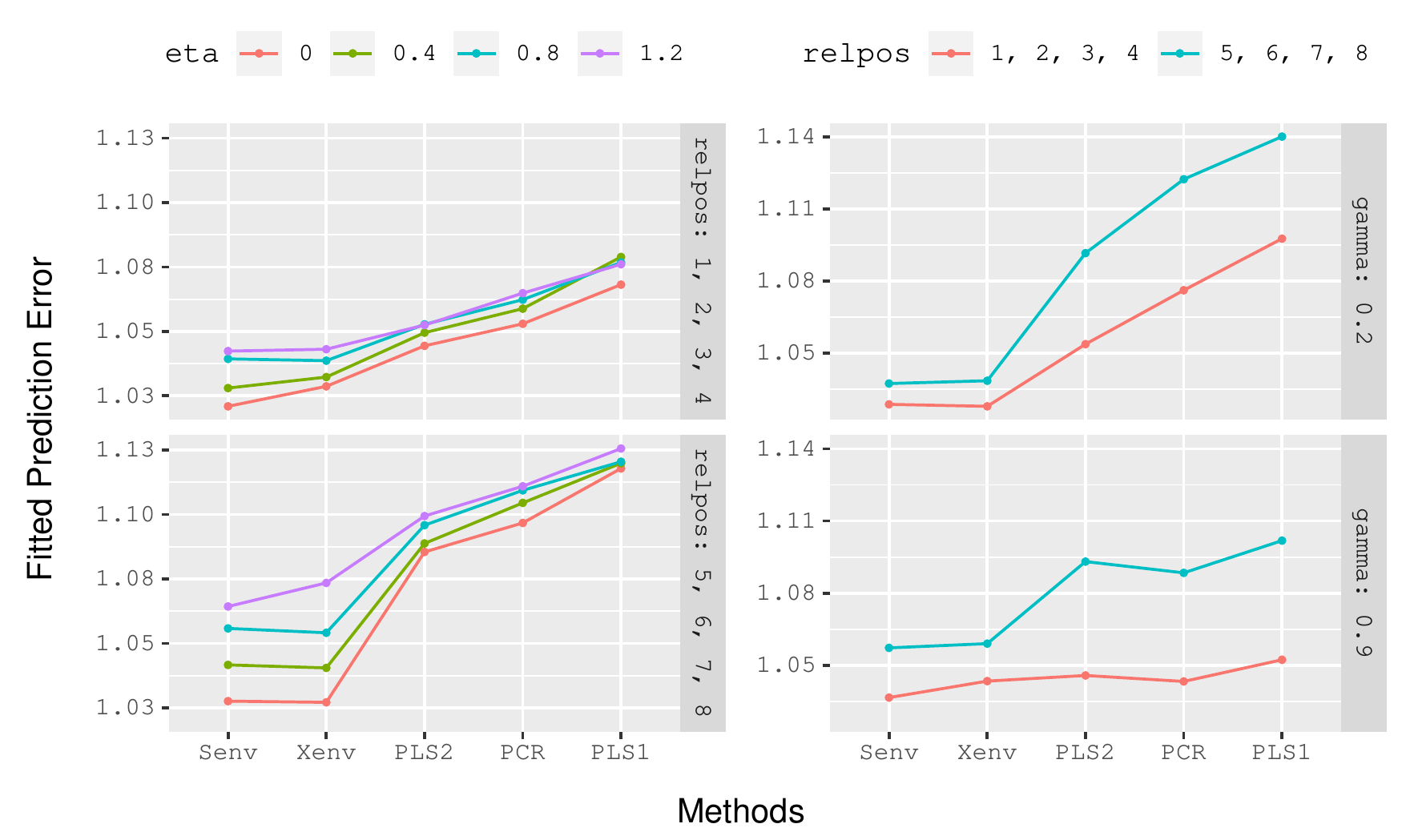} \caption{Effect plot of some interactions of the multivariate
linear model of prediction error}\label{fig:pred-eff-plots}
\end{figure}

The large difference in the prediction error for the envelope models in
Figure \ref{fig:pred-eff-plots} (left) is intensified when the position
of the relevant predictor is at 5, 6, 7, 8. The results also show that
the envelope methods are more sensitive to the levels of \texttt{eta}
than the rest of the methods. In the case of PCR and PLS, the difference
in the effect of levels of \texttt{eta} is small.

In Figure \ref{fig:pred-eff-plots} (right), we can see that the
multicollinearity (controlled by \texttt{gamma}) has affected all the
methods. However, envelope methods have better performance on low
multicollinearity, as opposed to high multicollinearity, and PCR, PLS1
and PLS2 are robust for high multicollinearity. Despite handling high
multicollinearity, these methods have higher prediction error in both
cases of multicollinearity than the envelope methods.

\subsection{Effect Analysis of Component
Model}\label{effect-analysis-of-component-model}

\begin{figure}[!htb]
\includegraphics[width=1\linewidth]{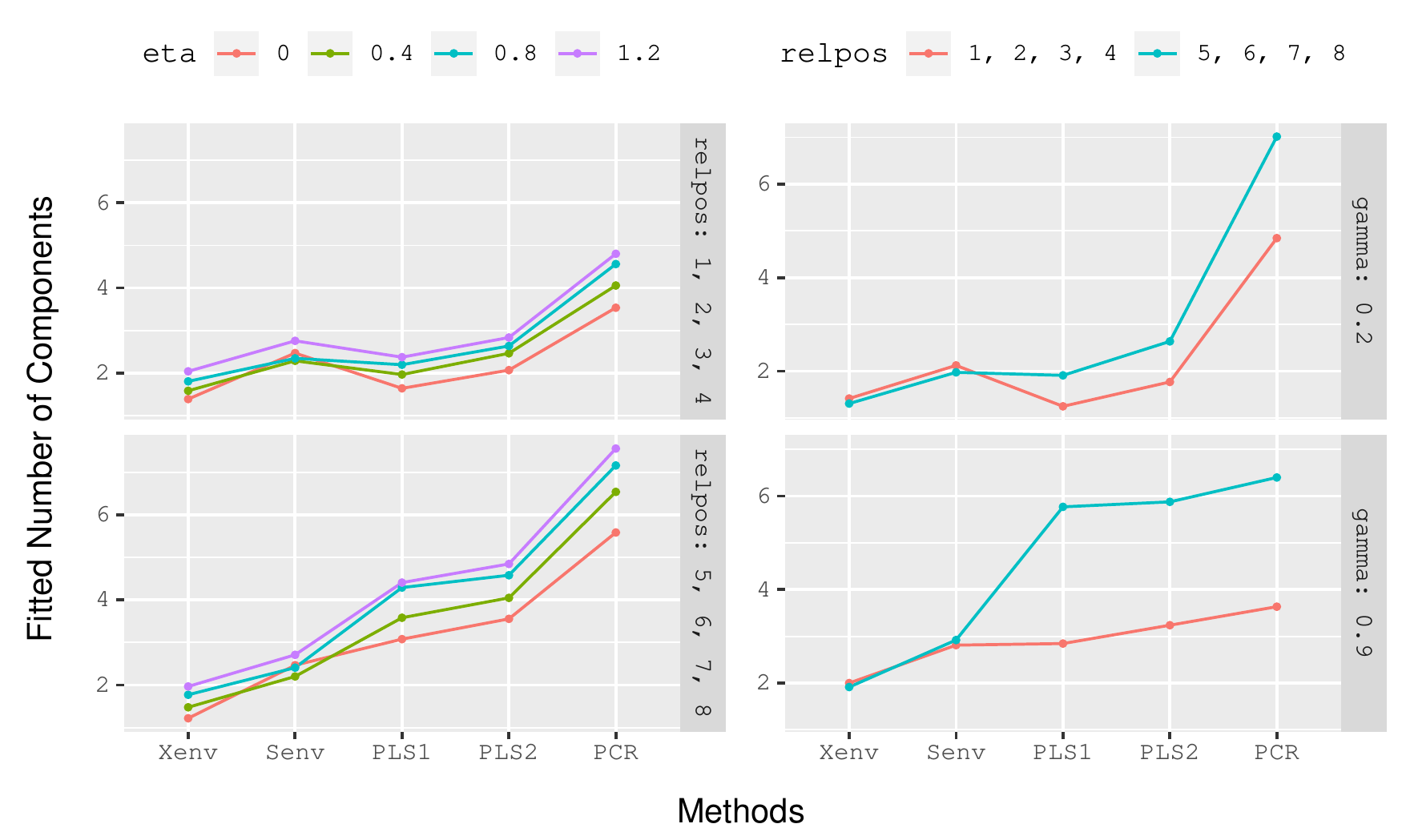} \caption{Effect plot of some interactions of the multivariate
linear model of the number of components to get minimum prediction error}\label{fig:comp-eff-plots}
\end{figure}

Unlike for prediction errors, Figure \ref{fig:comp-eff-plots} (left)
shows that the number of components used by the methods to obtain
minimum prediction error is less affected by the levels of \texttt{eta}.
All methods appear to use on average more components when eta increases.
Envelope methods are able to obtain minimum prediction error by using
components ranging from 1 to 3 in both the cases of \texttt{relpos}.
This value is much higher in the case of PCR as its prediction is based
only on the principal components of the predictor matrix. The number of
components used by this method ranges from 3 to 5 when relevant
components are at positions 1, 2, 3, 4 and 5 to 8 when relevant
components are at positions 5, 6, 7, 8.

When relevant components are at position 5, 6, 7, 8, the eigenvalues of
relevant predictors becomes smaller and responses are relatively
difficult to predict. This becomes more critical for high
multicollinearity cases. Figure \ref{fig:comp-eff-plots} (right) shows
that the envelope methods are less influenced by the level of
\texttt{relpos} and are particularly better in achieving minimum
prediction error using a fewer number of components than other methods.

\section{Examples}\label{examples}

In addition to the analysis with the simulated data, the following two
examples explore the prediction performance of the methods using real
datasets. Since both examples have wide predictor matrices, principal
components explaining 97.5\% of the variation in them are used for
envelope methods. The coefficients were transformed back after the
estimation.

\subsection{Raman spectra analysis of contents of polyunsaturated fatty
acids
(PUFA)}\label{raman-spectra-analysis-of-contents-of-polyunsaturated-fatty-acids-pufa}

This dataset contains 44 training samples and 25 test samples of fatty
acid information expressed as: a) percentage of total sample weight and
b) percentage of total fat content. The dataset is borrowed from
\citet{naes2013multi} where more information can be found. The samples
were analysed using Raman spectroscopy from which 1096 wavelength
variables were obtained as predictors. Raman spectroscopy provides
detailed chemical information from minor components in food. The aim of
this example is to compare how well the prediction methods that we have
considered are able to predict the contents of PUFA using these Raman
spectra.

\begin{figure}
\includegraphics[width=1\linewidth]{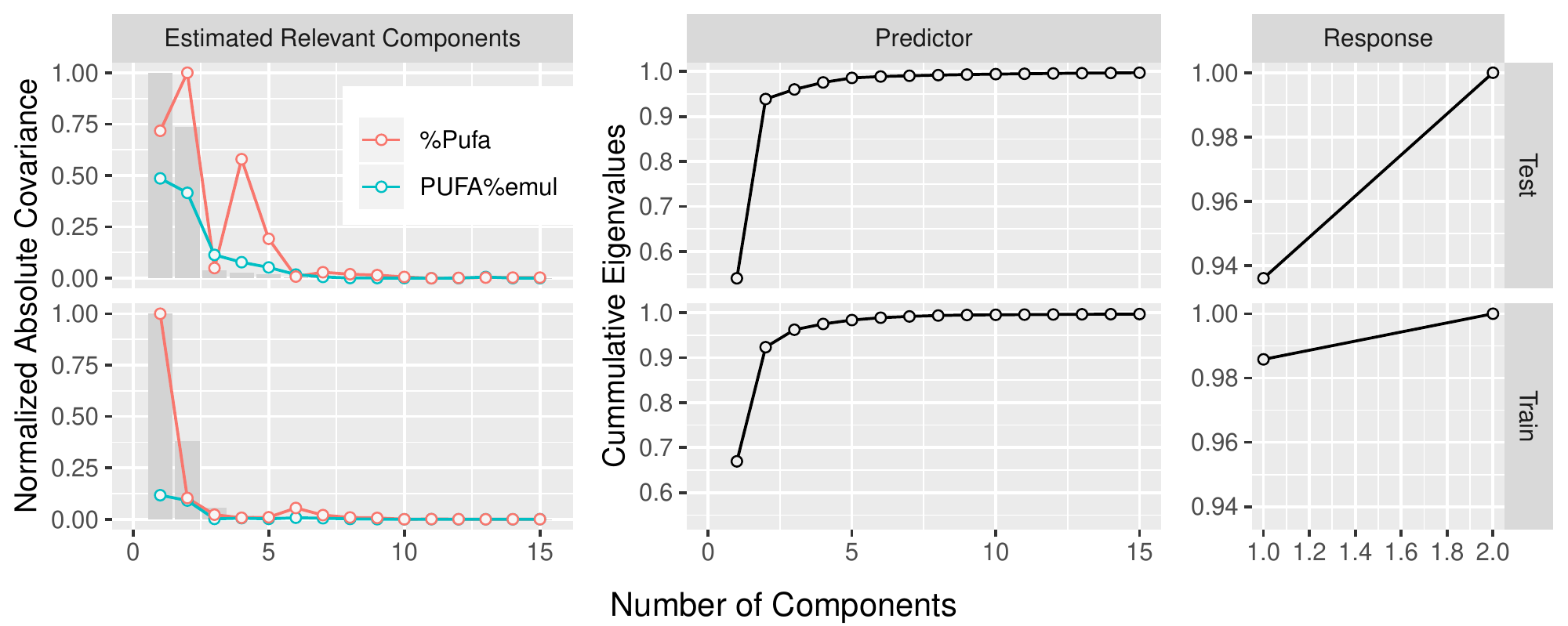} \caption{(Left) Bar represents the eigenvalues
corresponding to Raman Spectra. The points and line are the covariances
between response and the principal components of Raman Spectra. All the
values are normalized to scale from 0 to 1. (Middle) Cumulative sum of
eigenvalues corresponding to predictors. (Right) Cumulative sum of
eigenvalues corresponding to responses. The top and bottom row
corresponds to test and training datasets respectively.}\label{fig:ex1-cumulative-ues}
\end{figure}

Figure \ref{fig:ex1-cumulative-ues} (left) shows that the first few
predictor components are somewhat correlated with response variables. In
addition, the most variation in predictors is explained by less than
five components (middle). Further, the response variables are highly
correlated, suggesting that a single latent dimension explains most of
the variation (right). We may therefore also believe that the relevant
latent space in the response matrix is of dimension one. This resembles
the Design 19 (Figure \ref{fig:design-plot}) from our simulation.

\begin{figure}[!htb]
\includegraphics[width=1\linewidth]{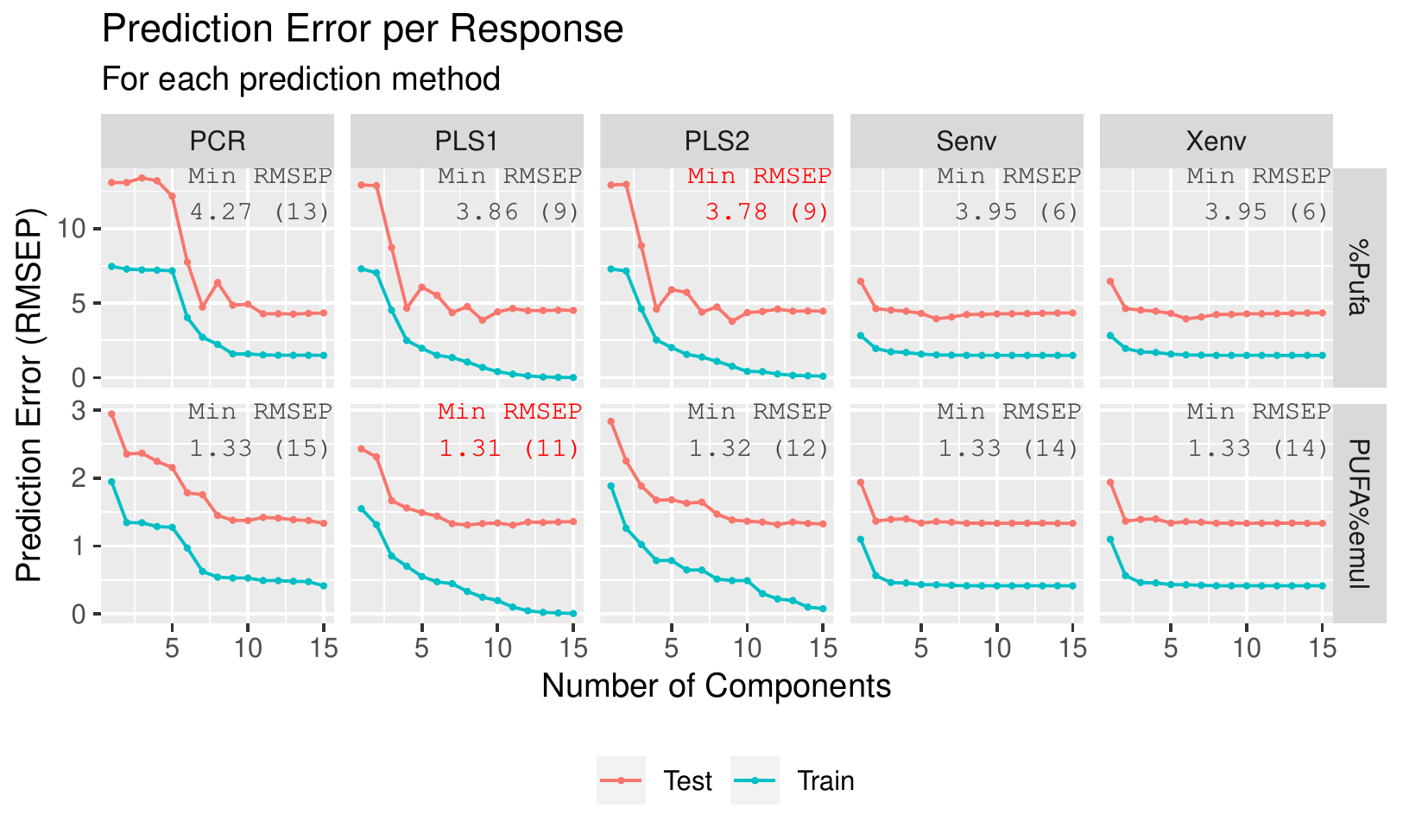} \caption{Prediction Error of different prediction methods using different number of components.}\label{fig:ex1-prediction-error}
\end{figure}

Using a range of components from 1 to 15, regression models were fitted
using each of the methods. The fitted models were used to predict the
test observation, and the root mean squared error of prediction (RMSEP)
was calculated. Figure \ref{fig:ex1-prediction-error} shows that PLS2
obtained a minimum prediction error of 3.783 using 9 components in the
case of response \%Pufa, while PLS1 obtained a minimum prediction error
of 1.308 using 11 components in the case of response PUFA\%emul.
However, the figure also shows that both envelope methods have reached
to almost minimum prediction error in fewer number of components. This
pattern is also visible in the simulation results (Figure
\ref{fig:comp-eff-plots}).

\subsection{Example-2: NIR spectra of biscuit
dough}\label{example-2-nir-spectra-of-biscuit-dough}

The dataset consists of 700 wavelengths of NIR spectra (1100--2498 nm in
steps of 2 nm) that were used as predictor variables. There are four
response variables corresponding to the yield percentages of (a) fat,
(b) sucrose, (c) flour and (d) water. The measurements were taken from
40 training observation of biscuit dough. A separate set of 32 samples
created and measured on different occasions were used as test
observations. The dataset is borrowed from \citet{indahl2005twist} where
further information can be obtained.

\begin{figure}
\includegraphics[width=1\linewidth]{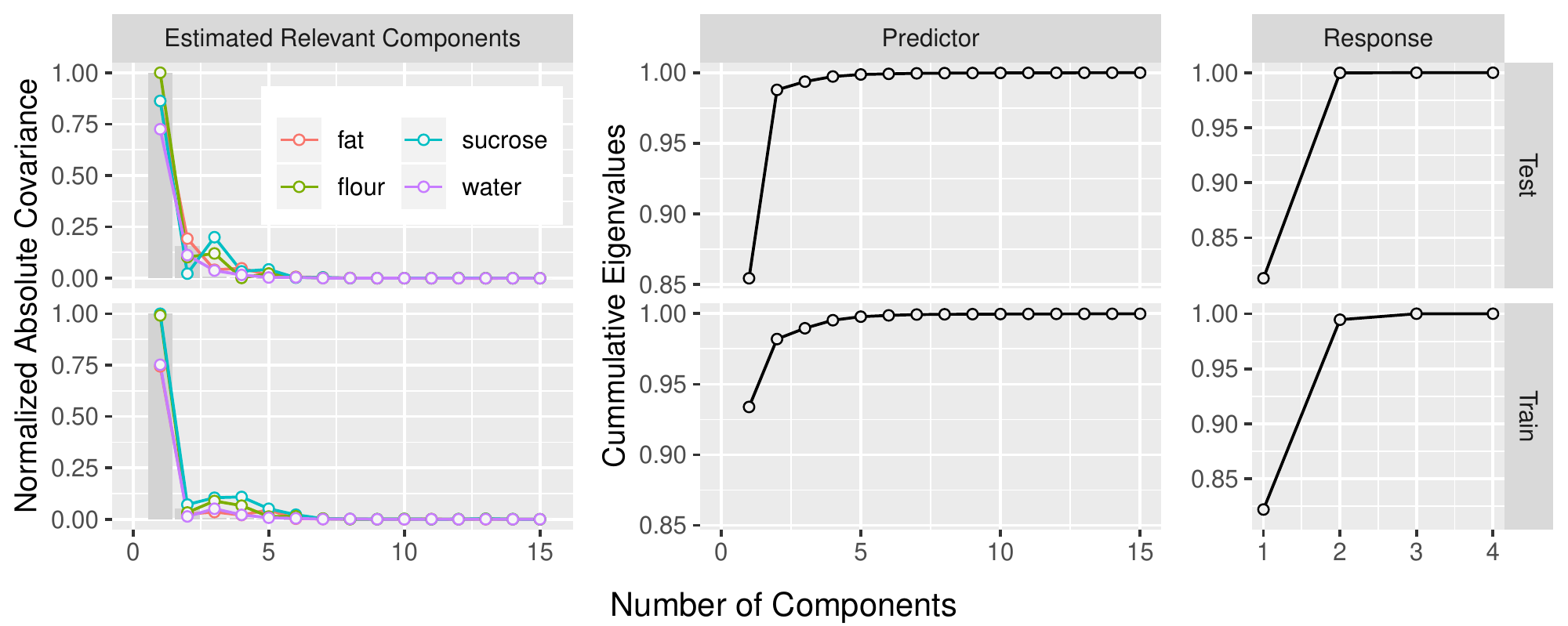} \caption{(Left) Bar represents the eigenvalues
corresponding to NIR Spectra. The points and line are the covariances
between response and the principal components of NIR Spectra. All the
values are normalized to scale from 0 to 1. (Middle) Cumulative sum of
eigenvalues corresponding to predictors. (Right) Cumulative sum of
eigenvalues corresponding to responses.}\label{fig:ex2-cumulative-eigenvalues}
\end{figure}

Figure \ref{fig:ex2-cumulative-eigenvalues} (left) shows that the first
predictor component has the largest variance and also has large
covariance with all response variables. The second component, however,
has larger variance (middle) than the succeeding components but has a
small covariance with all the responses, which indicates that the
component is less relevant for any of the responses. In addition, two
response components have explained most of the variation in response
variables (right). This structure is also somewhat similar to Design 19,
although it is uncertain whether the dimension of the relevant space in
the response matrix is larger than one.

\begin{figure}[!htb]
\includegraphics[width=1\linewidth]{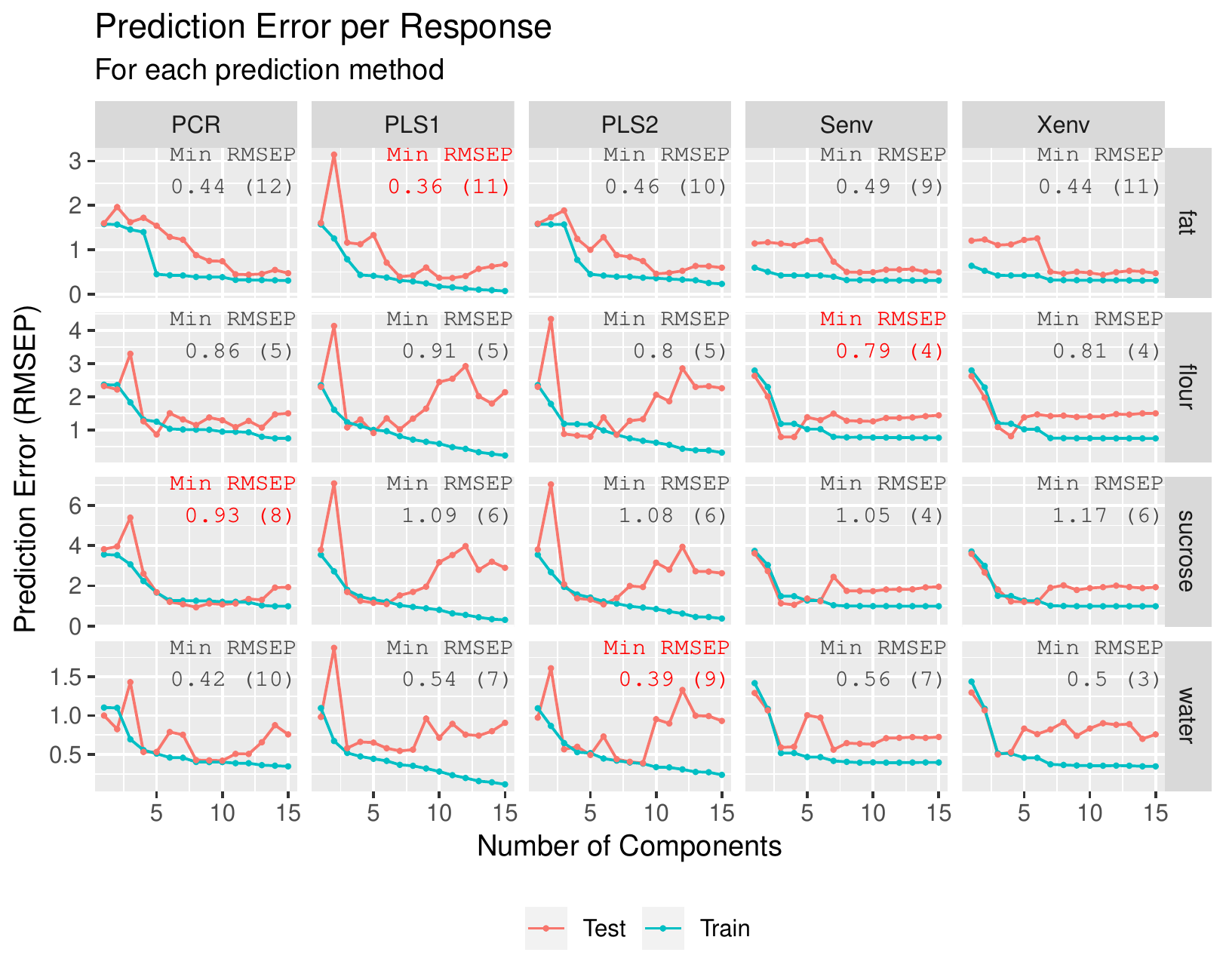} \caption{Prediction Error of different prediction methods using different number of components.}\label{fig:ex2-prediction-error}
\end{figure}

Figure \ref{fig:ex2-prediction-error} (corresponding to Figure
\ref{fig:ex1-prediction-error}) shows the root mean squared error for
both test and train prediction of the biscuit dough data. Here four
different methods have minimum test prediction error for the four
responses. As the structure of the data is similar to that of the first
example, the pattern in the prediction is also similar for all methods.

The prediction performance on the test data of the envelope methods
appears to be more stable compared to the PCR and PLS methods.
Furthermore, the envelope methods achieve good performance generally
using fewer components, which is in accordance with Figure
\ref{fig:comp-pca-hist-mthd-gamma-relpos}.

\section{Discussions and Conclusion}\label{discussions-and-conclusion}

Analysis using both simulated data and real data has shown that the
envelope methods are more stable, less influenced by \texttt{relpos} and
\texttt{gamma} and in general, performed better than PCR and PLS
methods. These methods are also found to be less dependent on the number
of components.

Since the facet in the Figures \ref{fig:pred-pca-hist-mthd-gamma-relpos}
and \ref{fig:comp-pca-hist-mthd-gamma-relpos} have their own scales,
despite having some large prediction errors seen at the right tail,
envelope methods still have a smaller prediction error and have used a
fewer number of components than the other methods.

Particularly in the case of the simultaneous envelope, since users can
specify the number of dimension for the response envelope, the method
can leverage the relevant space of response while PCR, PLS and Xenv are
constrained to play only on predictor space. \emph{Since the simulation
is based on a single latent component of the response variables, this
might have given some advantage for the simultaneous envelope.}

Furthermore, we have fixed the coefficient of determination (\(R^2\)) as
a constant throughout all the designs. Initial simulations (not shown)
indicated that low \(R^2\) affects all methods in a similar manner and
that the MANOVA is highly dominated by \(R^2\). Keeping the value of
\(R^2\) fixed has allowed us to analyze other factors properly.

Two clear comments can be made about the effect of correlation of
response on the prediction methods. The highly correlated response has
shown the highest prediction error in general and the effect is most
distinct in envelope methods. Since the envelope methods identify the
relevant space as the span of relevant eigenvectors, the methods are
able to obtain the minimum average prediction error by using a lesser
number of components for all levels of \texttt{eta}.

To our knowledge, the effect of correlation in the response on PCR and
PLS methods has been explored only to a limited extent. In this regards,
it is interesting to see that these methods have applied a large number
of components and returned a larger prediction error than envelope
methods in the case of highly correlated responses. To fully understand
the effect of \texttt{eta}, it is necessary to study the estimation
performance of these methods with different numbers of components.

In addition, since using principal components or actual variables as
predictors in envelope methods has shown similar results, we have used
principal components that have explained 97.5\% of the variation as
mentioned previously in the cases of envelope methods for the designs
where \(p>n\). As the envelope methods are based on MLE, this can be an
alternative way of using the methods in data with wide predictors. The
results from this study will help researchers to understand these
methods for their performance in various linear model data and encourage
them to use newly developed methods such as the envelopes.

Since this study has focused entirely on prediction performance, further
analysis of the estimative properties of these methods is required. A
study of estimation error and the performance of methods on the
non-optimal number of components can give a deeper understanding of
these methods.

A shiny application \citep{shiny} is available at
\url{http://therimalaya.shinyapps.io/Comparison} where all the results
related to this study can be visualized. In addition, a GitHub
repository at
\url{https://github.com/therimalaya/03-prediction-comparison} can be
used to reproduce this study.

\section{Acknowledgment}\label{acknowledgment}

We are grateful to Inge Helland for his inputs on this paper throughout
the period. His guidance on the envelope models and his review of the
paper helped us greatly. Our gratitude also goes to thank Kristian
Lillan, Ulf Indahl, Tormod Næs, Ingrid Måge and the team for providing
the data for analysis.

\hypertarget{refs}{}

\appendix

\renewcommand\refname{References}
\bibliography{ref-db.bib}

\end{document}